\documentclass[twocolumn,aps,floats,superscriptaddress,showpacs]{revtex4}

\usepackage{amsmath}
\usepackage{epsfig}
\usepackage{graphicx}
\usepackage{dcolumn}
\usepackage{bm}

\newcommand{\be}{\begin{equation}}
\newcommand{\ee}{\end{equation}}
\newcommand{\bea}{\begin{eqnarray}}
\newcommand{\eea}{\end{eqnarray}}

\newcommand{\Eq}[1]{Eq.\,(\ref{#1})}
\renewcommand{\dag}{^{\dagger}}

\newcommand{\la}{\left<}
\newcommand{\ra}{\right>}

\newcommand{\bdq}{b\dag_\mathbf{q}}
\newcommand{\bq}{b_\mathbf{q}}

\newcommand{\q}{\mathbf{q}}
\renewcommand{\r}{\mathbf{r}}
\newcommand{\R}{\mathbf{R}}
\newcommand{\Btl}{\tilde{B}_\lambda}

\begin{document}

\title{
Optical absorption in quantum dots:\\
Coupling to longitudinal optical phonons treated exactly}
\author{T. Stauber}
\affiliation{Instituto de Ciencia de Materiales de Madrid, CSIC,
Cantoblanco, E-28049 Madrid, Spain}
\author{R. Zimmermann}
\affiliation{Institut f\"ur Physik, Humboldt-Universit\"at zu Berlin,
Newtonstrasse 15, D-12489 Berlin, Germany}
\date{\today}
\begin{abstract}
Optical transitions in a semiconductor quantum dot are
theoretically investigated, with emphasis on the coupling to
longitudinal optical phonons, and including excitonic effects.
When limiting to a finite number of $m$ electron and $n$ hole
levels in the dot, the model can be solved exactly within
numerical accuracy. Crucial for this to work is the absence of
dispersion of the phonons. A suitable orthogonalization procedure
leaves only $m(m+1)/2+n(n+1)/2-2$ phonon modes to be coupled to
the electronic system. We calculate the linear optical
polarization following a delta pulse excitation, and by a
subsequent Fourier transformation the resulting optical
absorption. This strict result is compared with a frequently used
approximation modeling the absorption as a convolution between
spectral functions of electron and hole, which tends to
overestimate the effect of the phonon coupling. Numerical results
are given for two electron and three hole states in a quantum dot
made from the polar material CdSe. Parameter values are chosen
such that a quantum dot with a resonant sublevel distance can be
compared with a nonresonant one.
\end{abstract}
%

%
%
%
%
\maketitle
\section{Introduction}
Quantum dots (QDs) based on semiconductor structures have
received great attention for almost two decades now
\cite{Koch93}. Whereas early experiments were only able to
measure ensemble averages over many quantum dots, it is nowadays
possible to address single quantum dots individually
\cite{BCWB92}. Photoluminescence and more recently absorption
\cite{Guest} has been measured with high spatial resolution,
showing distinct lines related to individual dots (due to well
width fluctuations in a quantum well). Using
magneto-photoluminescence, detailed information on energy levels
and phonon coupling could be extracted for single QDs made from
II-VI semiconductor material \cite{FlissHen01}. This and related
experimental work has enormously stimulated investigations on the
interaction between the carriers on confinement levels and the
surrounding polarizable medium on a microscopic level. A proper
understanding of the related dephasing mechanisms might become
important in view of future quantum computation applications
based on semiconductor quantum dots \cite{Jacak05}.

Carrier relaxation and dephasing in quantum dots crucially
depends on the interaction with lattice vibrations (phonons). At
first sight, scattering with longitudinal-optical (LO) phonons
seems to be possible only if a sublevel spacing matches the LO
phonon energy $\hbar\omega_0$. In other (non-resonant) cases,
scattering is expected to be impossible (or at least strongly
reduced). This is the so-called phonon-bottleneck problem which
has been discussed intensively in the literature \cite{BB90}. The
argument relies on the application of Fermi's golden rule which
demands energy conservation in each individual process. However,
electron-phonon interaction in quantum confined systems
\cite{SMC87,VFB02} gives rise to a far more complicated picture
including: Non-Markovian scattering, formation of electron-phonon
bound states with level repulsion, and phonon satellites spaced
at multiples of $\hbar\omega_0$.

The coupling of quantum dot levels to acoustic phonons is
responsible for another set of features. Due to their dispersion,
the possible energy transfer covers a range from zero to a
maximum value (typically, a few meV) which is related to the dot
size. Consequently, phonon satellites appear here as broad bands
surrounding the zero-phonon line \cite{Besombes,Axt}. The
broadening of this line itself is a subject of intense research.
Similar to the phonon bottleneck problem mentioned above, an
advanced theory \cite{MZ04} predicts a finite broadening even in
cases where the next confined level is far away in energy
compared to the maximal possible energy transfer.

In the present work, we want to focus on the interaction with LO
phonons. This is relevant for quantum dots made from polar
material like CdSe, where the Fr\"ohlich interaction with LO
phonons is much stronger than the coupling to acoustic phonons.
Kr\'al et al. \cite{KK98} have dealt with the LO phonon
bottleneck problem and found within a standard self energy
approach (second order in the interaction) an appreciable
broadening of the levels which persists even outside the exact
resonance. It has further been argued that relaxation properties
can be obtained from a convolution of the spectral functions of
the discrete energy levels.

Aiming at a non-perturbative treatment, a big advantage is the
dispersionless nature of the LO phonons which allows a
numerically exact treatment. We have derived in our previous
paper Ref.\,\onlinecite{SZC00} an efficient scheme to calculate
eigenenergies and eigenvectors, which rests upon a unitary
transformation of the phonons into a new set where only a few
modes are coupled to the electronic degrees of freedom. We have
calculated the electron spectral function and shown that it
consists of a series of discrete delta functions. They are
distributed around the bare level energies and at multiples of
the LO energy (phonon satellites). Under resonance conditions,
the eigenenergies are still split in form of avoided level
crossing. This is reminiscent to the Rabi splitting in a
two-level system coupled to monochromatic photons (which formally
replace the phonons). The self consistent second Born
approximation for the self energy gives some gross features of
the spectrum but has broadened levels instead of the closely
spaced discrete lines in the exact calculation. Thus, we had
concluded that this approximation fails if phonons are coupled to
\emph{discrete} electronic levels. Recently, our numerically exact treatment was used to assess results obtained via the Davydov's canonical transformation \cite{Jacak03}.

Looking at the influence of phonon interaction on the
\textit{interband} transitions, one has to refer to a seminal
paper by Schmitt-Rink and coworkers \cite{SMC87}. They have shown
that the phonon interaction in general will increase with
confinement (i.e. with reducing the dot size). For the Fr\"ohlich
coupling to LO phonons, however, things are more subtle since
here the difference in charge distribution between initial
(valence band sublevel) and final state (conduction band
sublevel) enters. Now, under strong confinement, the sublevel
wave functions are getting more similar. Therefore, the optical
transition is accompanied with practical no change in local
charge distribution, and the net LO phonon coupling is
drastically reduced.

In Ref.\,\cite{KK98}, it has been claimed that the interband
absorption spectrum can be taken as a convolution of the
one-particle spectral functions for the electron and the hole. In
the present work which extends Ref.\,\cite{SZC00} to a two-band
many-level situation, and allows to calculate the linear optical
response exactly, we will demonstrate the failure of this
convolution approach. Indeed, it is missing the correct charge
redistribution.

Absorption in QDs was first discussed in Ref. \cite{Fom98} with
emphasis on the non-adiabatic electron-phonon coupling, labeled
as phonon-assisted transitions in Ref. \cite{SZC00}. In Ref.
\cite{Jac03}, a detailed comparison between features in QDs and
phenomena from quantum optics was presented, again focusing on
the electron-phonon interaction. Recently, the absorption
spectrum of individual QDs has been discussed for intra-band
\cite{LOH05} as well as for inter-band transitions \cite{VAM04}.
In the latter work, the valence band levels were treated without
phonon interaction - which leads to an even simplified form of
the convolution approach. However, the resonance condition
(matching a sublevel distance by $\hbar\omega_0$) can be much
easier obtained in the valence band since the sublevel spacing is
here smaller than in the conduction band.

The paper is organized as follows. In section \ref{TheModel}, the
model and the Hamiltonian are presented, which includes both
carrier-phonon and carrier-carrier (Coulomb) interaction. In
section \ref{GramSchmidt}, we define and perform the unitary
transformation which reduces the number of Bosonic modes coupled
to the Fermionic states. This works if the phonons have no
dispersion, and makes a numerical diagonalization of the
Hamiltonian feasible. Equations for the linear polarization in
terms of exact eigenstates are derived in section
\ref{SecAbsorption}, and the approximate convolution of electron
and hole spectral function is given as well. In section
\ref{Numerics}, numerical results for two prototype quantum dots
made from the polar material CdSe are presented and discussed.
Several details on the transformations and a list of material
parameters are given in the Appendix.
%
%
%
%

\section{The model}\label{TheModel}

We start with the standard Hamiltonian which couples the band
states in a semiconductor to the lattice displacement. For the
quantum dot, the confined electronic states in the conduction
(valence) band are given by Fermionic creation operators $c\dag_j
\,(v\dag_j)$. The lattice vibrations are taken as longitudinal
optical phonons without dispersion, $\hbar\omega_0$, and
represented by Bosonic operators $\bq\dag$:
\bea \label{HamCV} H & = & \sum_{\q}\hbar\omega_0 \bdq \bq
 + \sum_j \left(\epsilon_j^c c_j\dag c_j \, + \,
 \epsilon_j^v v_j\dag v_j\right)\\
& + & \sum_{\q jl}\left(\bq + b_{-\q}\dag \right)\left(M_{\q jl}^c
c_j\dag c_l + M_{\q jl}^v v_j\dag v_l\right) \, .\nonumber \eea
The coupling matrix elements $M_{\q jl}^a \, (a=c,v)$ stem from
the Fr\"{o}hlich interaction applied to the confinement wave
functions (see Appendix \ref{AppCoupling} for details). Note that
there is no phonon coupling between valence and conduction band
states since the energy gap is far greater than the LO phonon
energy, while the distance between confined levels in one band
$\epsilon_j^a - \epsilon_l^a \, (a = c,v)$ may be well in the
range of the phonon energy. Electron spin is not included here
since spin relaxation is a slow process compared to the
spin-conserving electron-phonon interaction. Similarly,
phonon-assisted transitions into the continuum of wetting layer
states are not considered. For simplicity, we restrict ourselves
to the interaction with bulk phonon modes, leaving the more
precise picture of confined and interface modes for future
investigations \cite{GKFNZGT04}.

For treating optical transitions between the filled valence band
and the empty conduction band, it is convenient to switch from the
conduction-valence-band description used in \Eq{HamCV} to the
electron-hole picture. This is accomplished by the replacement
\be c_j\rightarrow e_j,\quad v_j\rightarrow h_j\dag \, . \ee
Using $v_j\dag v_l=\delta_{j,l}-h_l\dag h_j$ the Hamiltonian  is
rewritten as
\bea \label{HamEH} H & = & \sum_\q \hbar\omega_0 \bdq \bq \, + \,
\sum_j \left(\epsilon_j^ce_j\dag e_j -\epsilon_j^vh_j\dag h_j\right)\\
& +  & \sum_{\q jl}\left(\bq + b_{-\q}\dag \right) \left(M_{\q
jl}^c e_j\dag e_l -
M_{\q lj}^v h_j\dag h_l\right) \\
& + & \sum_j\epsilon_j^v \, +\, \sum_{\q j}\left(\bq +
b_{-\q}\dag\right) M_{\q jj}^v \, . \nonumber \eea
Note that the phonon interaction now carries a negative sign for
the hole states compared to the electron states, which can be
traced back to the different charge sign of the excitations. In
Appendix \ref{AppEHTrafo} we show that the last line can be
dropped by renormalizing the ground state energy.

As we are focusing on the linear response, the Coulomb
interaction leads to the formation of a single exciton only. The
relevant interaction term is
\bea \label{HamCou} H^{C} = -\sum_{klij} v_{klij} e_k\dag h_l\dag
h_je_i\,,\eea
and has to be added to \Eq{HamEH}. $v_{klij}$ denotes the Coulomb
matrix element between sublevel states (for details, see appendix
\ref{AppCoupling}). Under strong confinement conditions
considered here, the confinement states are only reshaped
marginally \cite{ZimEdinb}, and the Coulomb interaction leads to
almost rigid shifts of the transition energies (exciton binding
energies). However, non-diagonal Coulomb couplings have to be
included since they can be of similar order as the phonon
couplings, and modify the oscillator strengths as well.
%
%
%

\section{Reducing the phonon subspace}\label{GramSchmidt}

The phonon coupling in \Eq{HamEH} involves only certain
combinations of phonon operators, e.g. $\sum_\q M_{\q jl}^c \,
\bq$. If we consider a finite number (say $m$) of electron
sublevels in the dot, and $n$ hole sublevels, there are
$m^2+n^2$ such combinations (as argued above, phonon-assisted
transitions between conduction and valence bands are absent).
Since we are dealing exclusively with confined states, the
confinement wave functions can be chosen to be \emph{real}, and
the symmetry $M_{\q lj}^a = M_{\q jl}^a $ holds. This reduces the
linear independent combinations to $m(m+1)/2+n(n+1)/2$.

A further reduction can be achieved since in the present
Hamiltonian \Eq{HamEH}, the number of electrons ($N_e$) and of
holes ($N_h$) are conserved quantities. Therefore, two (diagonal)
Fermion pairs in the interaction can be expressed by the remaining
ones. We choose
\be e\dag_m \, e_m = N_e - \sum_{j=1}^{m-1} e\dag_j \,e_j \, ,
\quad h\dag_n \, h_n = N_h - \sum_{j=1}^{n-1} h\dag_j \,h_j \,
,\ee
which gives additions to all the other diagonal coupling terms,
and a $c$ number remainder. In order to shorten the subsequent
writing, we introduce pair indices $\lambda = 1, \dots ,
(N-2)$ which combine either two electron sublevels (e:$jl$)
or two hole sublevels (h:$jl$), with $j \le l$. Caring for the
different signs in the interaction of \Eq{HamEH}, we define in the
conduction band
\be M_{\q\lambda}  =  M^c_{\q\,jl} - \delta_{jl} M^c_{\q\,mm} \,
,\ee and in the valence band \be M_{\q\lambda}  =  -M^v_{\q\,lj} +
\delta_{jl} M^v_{\q\,nn} \, .\ee
For treating the $c$ number part properly, we have to introduce
one further index $\lambda = \kappa \equiv N-1$, with the
matrix element
\be M_{\q\,\kappa} = N_e M_{\q\,mm}^c \, - \, N_h M_{\q\,nn}^v
\,.\ee
Now, the interaction term of \Eq{HamEH} reads
\be \label{HamInt} \sum_{\lambda=1}^{\kappa-1} \left( A_\lambda \,
+ \, A\dag_\lambda \right) \left(c\dag\,c \right)_\lambda \,
  + \, \left( A_\kappa \, + \,
A\dag_\kappa \right) \ee
with a shorthand writing for the Fermionic operators. Note that
for the nondiagonal terms $j \neq l$, we have to set e.g.
\be \label{OpSum} (c\dag c)_\lambda = e\dag_j e_l + e\dag_l e_j \,
. \ee
The combinations of phonon operators entering \Eq{HamInt} are
\be A_\lambda = \sum_\q M_{\q\lambda} \bq  \quad \left(\lambda =
1\dots \kappa \right) \, ,\ee
but the $A_\lambda$ do not form an orthonormal set. As detailed in
Appendix \ref{AppGramSchmidt}, we apply the Gram-Schmidt
orthonormalization scheme to generate new phonon operators
$B_\lambda$ which are properly orthogonalized. The transformation
is a linear one,
\be A_\lambda = \sum_{\alpha=1}^\lambda I_{\lambda\alpha} B_\alpha
\, ,\ee
and leads to the transformed Hamiltonian
\bea \label{HamB} H &=& \sum_\lambda \hbar\omega_0 B\dag_\lambda
\, B_\lambda \, + \, \sum_j \left( \epsilon^c_j \, e\dag_j
\, e_j \, - \, \epsilon^v_j \, h\dag_j \, h_j\right) \nonumber \\
& + & \sum_{\lambda=1}^{\kappa-1} \sum_{\alpha=1}^\lambda
   \left( I_{\lambda \alpha} B_\alpha \, + \, I^*_{\lambda \alpha}
     B\dag_\alpha \right) \left(c\dag c\right)_\lambda \\
& + & \sum_{\alpha=1}^{\kappa} \left(I_{\kappa \alpha} B_\alpha
   \, + \, I^*_{\kappa \alpha} B\dag_\alpha \right) \,. \nonumber \eea
To arrive at the standard diagonal form of the free phonon
part, it was essential that the phonons have no dispersion.
Otherwise, the subset of $B_\lambda$ would still mix with the
remaining phonon operators. These remaining degrees of freedom
only contribute to the free phonon energy and are omitted from
\Eq{HamB}.

A careful inspection of \Eq{HamB} shows that the Bosonic operator
$B_\kappa$ only appears as last element in the last line of
\Eq{HamB}, and in the free phonon part. It is therefore decoupled
from the Hamiltonian and allows an independent solution in terms
of a shifted oscillator,
\be {\cal B}_\kappa = B_\kappa + I_{\kappa\kappa} /\hbar\omega_0
\, . \ee
Therefore, only
\be \label{ModeNumber} N = \frac{m(m+1)}{2} \, +\,
\frac{n(n+1)}{2} - 2 \ee
new modes are coupled to the Fermionic Hilbert space.

Assuming a symmetric dot shape, the confinement functions are
either even or odd, and the matrix elements $M_{\q jl}^a$ have a
definite parity in $\q$, too. Consequently, the compound matrix
elements $M_{\lambda\alpha}$ defined in \Eq{Mlambdaalpha} vanish
if $\lambda$ and $\alpha$ refer to different parity. Therefore,
the tridiagonal matrix $K$ (and $I$ as well) has a block
structure, allowing the Gram-Schmidt procedure to work in each
block independently. Note that the Coulomb matrix elements have
an equivalent parity, and therefore do not mix these blocks. For
certain confinement potentials like the harmonic potential, the
number of phonon modes can be reduced even further (see appendix
\ref{AppCoupling}).
%
%
%
%
\begin{figure}[t]
\begin{center}
\includegraphics*[width=1.0in,angle=-90]{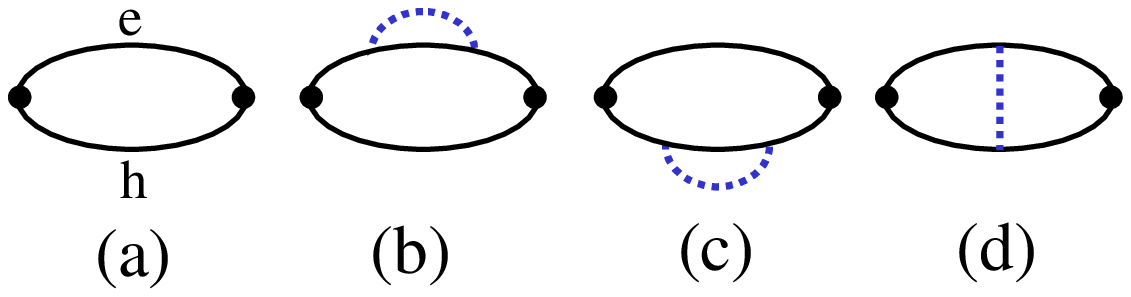}
\caption{\label{Diagram}Diagrammatic expansion of the optical
polarization up to first order in the electron-phonon interaction.
Lines denote the electron (e) and hole (h) propagator, the dotted
line the phonon propagator, and the large dot stands for the
optical dipole matrix element. The zeroth-order diagram (a)
describes interband transitions between confined states unaffected
by the phonon interaction. In the convolution approach, only
self-energy type diagrams (b, c) are considered, while in the full
evaluation, vertex type diagrams (d) are included as well.}
\end{center}
\end{figure}

\section{Linear optical response}\label{SecAbsorption}

In this section, we want to contrast the direct evaluation of the
time-dependent linear polarization and the absorption spectrum
with the so-called \emph{convolution approach}: Here, only the
spectral functions of the electron and hole states are
calculated, and the absorption is taken as a convolution of both
quantities. Before deriving the corresponding formal expressions,
let us point out the main difference: For the direct evaluation,
we start from the electron-hole vacuum ($N_e=0, N_h = 0$) as
initial state and end up in the subspace of one electron-hole
pair ($N_e=1, \, N_h = 1$). For the convolution approach, quite
different subspaces are invoked, namely $N_e=0, N_h = 0
\rightarrow N_e=1, N_h = 0$ for the electron spectral function,
and $N_e=0,\, N_h=0 \rightarrow N_e=0,\, N_h=1$ for the hole one.
In diagram language, any process which has a phonon correlation
between electron and hole levels is discarded in the convolution
approach. In Fig.\,\ref{Diagram}, we display the relevant first
order diagrams. The vertex diagram Fig.\,1d is the first one
describing interband phonon correlations. In particular under
strong confinement conditions, it compensates to a large extent
the self-energy type diagrams (Fig.\,1b, c). Only the latter are
kept in the convolution approach and we expect much to strong
phonon satellites in this approximate treatment. Formally, in the
convolution approach, the matrix elements appear as $(M^c)^2$ or
$(M^v)^2$, while the full version includes the vertex corrections
containing $M_c M_v$ as well. For level-diagonal matrix elements,
this combines into $(M^c-M^v)^2$. The appearance of matrix
element differences can be seen already in \Eq{HamEH}. Therefore,
the near cancellation of changes in the charge distribution
\cite{SMC87} can only be achieved when treating self energy and
vertex diagrams on an equal footing. Note that excitonic effects
are completely absent within the convolution approach.

\subsection{Direct evaluation}
The coupling to the light is described within the dipole
approximation for interband transitions, adding
\be\label{IncomingField} E(t)\sum_{ij}(\mu_{ij}\,c_i\dag v_j \, +
\, h.c.) \ee
as a perturbation to the Hamiltonian. The dipole matrix elements
are given by an integral over the confinement functions,
\be\label{DipoleDef} \mu_{ij}=\mu_{cv}\int d^3 r\,\psi_i^e({\bf
r})\,\psi_j^h({\bf r}) \, ,\ee
having as prefactor the dipole moment between the valence and
conduction band, $\mu_{cv}$.

The linear optical response follows from the polarization after a
unit-area delta pulse at $t=0$ and is given as
dipole-dipole-correlation function. Expressed via the electron and
hole operators, we have for $t \ge 0$
\be P(t) = i\sum_{ij,kl}\mu_{kl}^*\,\mu_{ij}\langle
h_l(t)\,e_k(t)\,e_i\dag(0)\,h_j\dag(0)\rangle \ee
with the time dependence of the operators in the Heisenberg
picture. The expectation value is shorthand writing for the
statistical sum over initial states which contain no electron-hole
pairs, and a thermal distribution of phonon mode occupations
$n_\lambda = 0, 1, \dots \infty$. This is denoted by
$|n_\lambda,0\rangle$, with total energy
\be E_0 = \hbar\omega_0\sum_\lambda n_\lambda \, . \ee
Then, we proceed with
\be P(t)\propto i\sum_{n_\lambda}e^{-\beta E_0}
\sum_{ij,kl}\mu_{kl}^*\mu_{ij}\langle n_\lambda,0|
h_l(t)e_k(t)|n_\lambda,ij\rangle \, . \ee
In practice, the independent summation over phonon occupations
$n_\lambda$ can be restricted to a maximum number which depends on
coupling strength and temperature, $\beta=1/k_BT$. For simplicity,
we have omitted a prefactor which ensures the proper normalization
of the statistical sum.

While the zero-pair states $|n_\lambda, 0\rangle$ diagonalize the
zero-pair Hamiltonian properly, we need to look for the one-pair
states from the eigenvalue problem
\be \label{Eigenvalue1} \left(H + H^C\right)_1|\phi\rangle =
E_\phi|\phi\rangle \, . \ee
Due to the reduction of the number of Bosonic modes, this can be
solved numerically exact by expanding into the noninteracting
one-pair basis $|n_\lambda,ij\rangle$.

Plugging all time dependencies together, we obtain
\bea \label{Poft} P(t)&=& i \sum_{n_\lambda}e^{-\beta
E_0}\sum_{ij,kl,\phi}
\mu_{kl} \,\mu_{ij} \\
&\times& e^{i(E_0-E_\phi)t/\hbar}\langle
n_\lambda,kl|\phi\rangle\langle\phi|n_\lambda,ij\rangle \, .
\nonumber \eea
Starting with the initial value
\be P(t=0) = i \sum_{n_\lambda} e^{-\beta E_0} \sum_{kl}
 |\mu_{kl}|^2\, ,\ee
the polarization evolves in time as a sum over many individual
oscillations. Although each of these terms does not have a
damping, in the sum a general decay can be observed
(quasi-dephasing), as shown in Sec.\,\ref{Numerics}.

The imaginary part of the Fourier transformation of the
polarization function yields the absorption spectrum,
\bea \label{DefAbsorption} \alpha(\omega)& = &
\text{Im}\int_{-\infty}^\infty dt \, P(t) \, e^{i\omega t} \\
& \propto& \sum_{n_\lambda} e^{-\beta E_0} \sum_{ij,kl,\phi}
\mu_{kl}^*\mu_{ij} \nonumber \\
&\times &\langle n_\lambda,kl| \phi\rangle
\langle\phi|n_\lambda,ij \rangle \delta(E_\phi-E_0-\hbar\omega) \,
. \eea
Let us point out that the exact absorption spectrum consists of
delta peaks only. This was clear from the beginning since a finite
perturbation cannot change the character of the spectrum of the
unperturbed system. Since we started from {\em dispersionless}
Bosonic modes, the discrete electronic spectrum cannot be altered
by the non-zero electron-phonon interaction \cite{ReSi75}.

\subsection{Convolution approach}
It has been argued in the literature \cite{KK98}, that the
convolution of electron and hole spectral functions may give a
reasonable approximation to the absorption spectrum,
\be \label{AbsConvol} \alpha^{con}(\omega)\propto \int d\omega'
\sum_{ij,kl}\mu_{kl}^*\,\mu_{ij}\,A_{ki}^e(\omega')\,
A_{lj}^h(\omega-\omega') \, . \ee
Indeed, avoiding the evaluation of the two-particle (i.e. dipole-dipole)
correlation functions would be an important reduction of the
numerical labor, since the spectral function is a genuine
one-particle function.
 We concentrate on
the electron spectral function $A_{ki}^e(\omega)$ which is defined
by
\be \label{SpectralDef} A_{ki}^e(\omega) = \frac{1}{\pi} \,
\text{Re} \int_{0}^{\infty} dt \, e^{i\omega t} \, \langle e_k(t)
e\dag_i(0)\rangle \, .\ee
Here, we need the exact eigenstates of the Hilbert space with
$N_e=1,\,N_h=0$ which will be denoted by $H_e|\chi\rangle =
E_\chi|\chi\rangle$. Similar arguments as used in deriving
\Eq{Poft} lead to
\bea \label{SpectralExpl} A_{ki}^e(\omega) & \propto
&\sum_{n_\lambda}
e^{-\beta E_0} \\
&\times& \sum_{\chi}  \langle n_\lambda,k|\chi\rangle
\langle\chi|n_\lambda,i\rangle \,\delta(E_\chi - E_0 -\hbar\omega)
\, , \nonumber \eea
where $|n_\lambda, k\rangle$ is the (non-interacting)
one-electron basis. Similar equations hold for the hole spectral
function. Due to the assumed dot symmetry, the only off-diagonal
spectral function will be the hole spectral function related to
h:13.

%
%
%
%
\section{Numerical results}\label{Numerics}
\begin{figure}[t]
\begin{center}
\includegraphics*[width=2.6in,angle=-90]{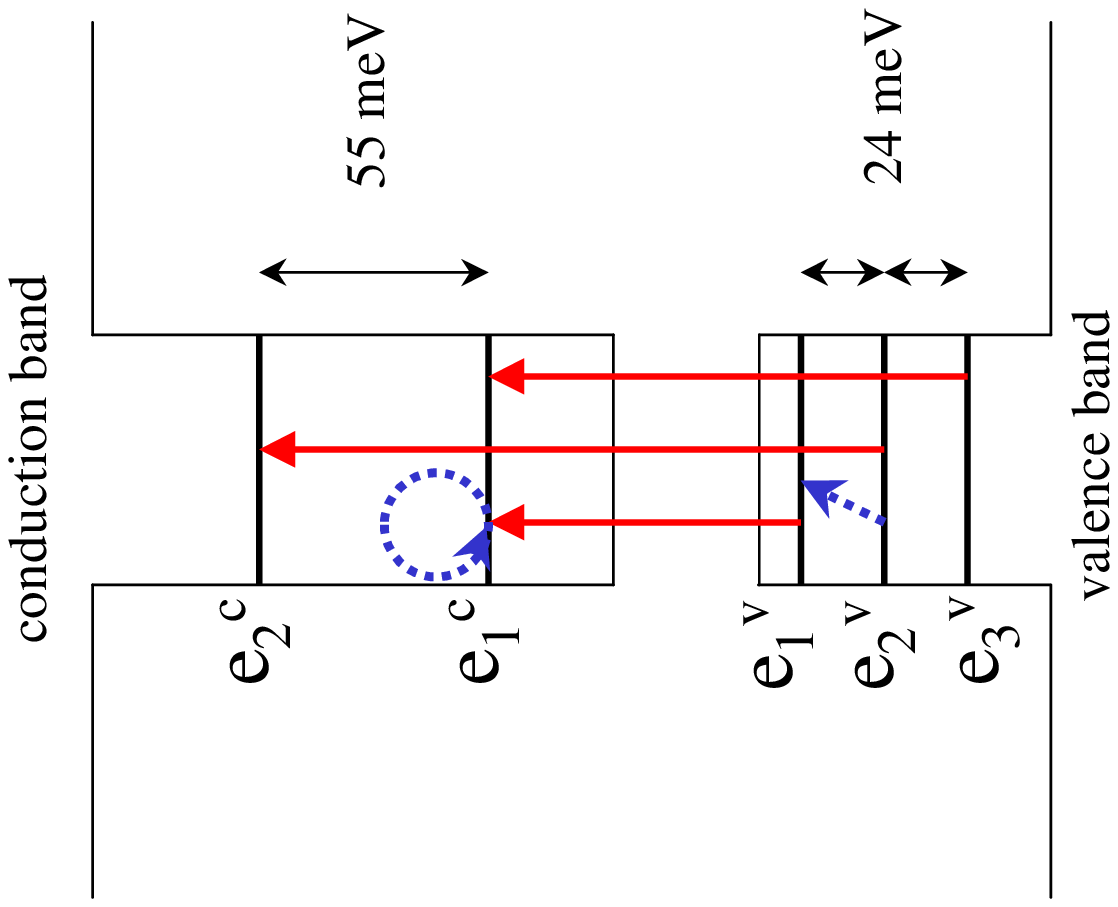}
\caption{\label{Levels}Schematic band diagram with sublevel
spacings corresponding to the resonant quantum dot. Vertical
arrows - optical transitions, dotted lines - phonon processes. For
further explanation, see text.}
\end{center}
\end{figure}

We consider a prototype quantum dot of ellipsoidal shape with
CdSe as the semiconductor material. The relevant LO phonon energy
is $\hbar\omega_0 = 24\,$meV. The larger mass of the holes
results in a smaller level distance in the valence band. In order
to cover a comparable energy range in the conduction and the
valence band, we take into account the two lowest confined states
in the conduction band ($m=2$) and the three lowest confined
states in the valence band ($n = 3$). In accordance with
\Eq{ModeNumber}, we then have seven new phonon modes.

According to what has been said before, there is the odd parity
group ($\lambda$ = e:12, h:12, h:23) and the even parity group
($\lambda$ = e:11, h:11, h:22). The special mode $\lambda=\kappa
\equiv 7$ refers to the even parity group. Consequently, $K$ and
$I$ decompose into a three-dimensional block for both parity
groups. Due to the harmonic potential, no extra Bosonic mode is
assigned to the transition h:13 which can be expressed by other
Bosonic modes (see Appendix \ref{AppCoupling}).

For characterizing the different dot sizes we define the first
sublevel distance in each band (without phonon interaction and
exciton effects) as
\be \Delta^a\equiv|\epsilon_2^a - \epsilon_1^a| \quad (a = c, v)
\, . \ee
In the sequel, we calculate the linear optical properties for two
selected dot sizes: The \emph{resonant quantum dot} has perfect
resonance between phonon energy and sublevel distance in the
valence band, $\Delta^v=\hbar\omega_0=24\,$meV, and consequently
$\Delta^c=55\,$meV. In Fig.\,\ref{Levels}, the confinement levels
for the resonant QD are shown schematically. Vertical arrows mark
the nonzero optical transitions. The dotted lines denote phonon
processes. In the valence band, this is resonant with the
sublevel distance (real transition, driven by $M_{\q 12}^v$). In
the conduction band, a virtual process involving the diagonal
matrix element $M_{\q 11}^c$ is depicted.

The second choice is backed by experimentally measured level
distances from Refs.\,\cite{FlissHen01,Akimov04} and comes out to
be a \emph{non-resonant quantum dot} with $\Delta^v=35\,$meV and
$\Delta^c=80\,$meV. More details on the extraction of size
parameters are given in Appendix \ref{AppParameters}.

\subsection{Direct evaluation}

To determine the full linear response, we need to solve the
eigenvalue problem for the one-pair subspace \Eq{Eigenvalue1}. As
mentioned at the end of Sec.\,\ref{TheModel}, it will consist of
six Fermionic pair states which are coupled to six Bosonic modes.
We truncate the Hilbert space by allowing the sum of occupation
numbers in all six Bosonic modes not to exceed seven. This gives
1716 different Bosonic states using the binomial coefficient (13
over 6). For the Fermionic states $\left |jl\ra$, we can exploit
an additional parity symmetry. Thus, in total, two matrices with
$5148\times5148$ elements each have to be diagonalized. At
$T=77\,$K, we have $\beta\hbar\omega_0 = 3.62$, which tells us
that LO phonon emission processes are dominant. By checking the
convergence, we found that the first 1000 eigenvalues are
sufficient to describe the spectra properly.

In dealing with the special phonon mode $B_7$, we have to evaluate
the matrix elements between the unshifted oscillator (in the
initial state) and the shifted one (in the final state),
\bea
&& \la n,0|n',s\ra = \\
&& e^{-s^2/2}  \sum_{k=0}^{\min(n,n')} \frac{(-1)^{k+n} \sqrt{n!
n'!}}{k!\,(n-k)!\,(n'-k)!} s^{n+n'-2k} \, .\nonumber \eea
The shift parameter is given as second argument in the state, here
$s = I_{77}/\hbar\omega_0$. In the polarization, this part gives
rise to an additional factor of
\be \label{Special} \sum_{n_7,n'_7} e^{-\beta n_7 \hbar\omega_0}
\, e^{i(n_7 - n'_7 - s^2)\omega_0 t} \,\left|\la n_7,0|n'_7,s \ra
\right|^2 \, . \ee
Note the appearance of $s^2\omega_0$ as a correction to the final state
energy. For the absorption spectrum, first the spectrum without
the special mode is calculated, and subsequently spectrally
displaced and added up, as the Fourier transform of \Eq{Special}
dictates.
\begin{figure}[t]
\begin{center}
\includegraphics*[width=2.8in,angle=0]{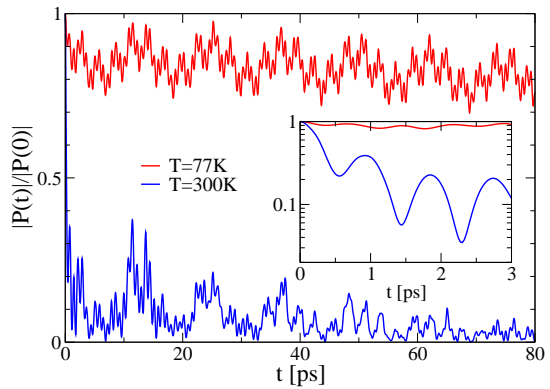}
\caption{\label{Polaris}Temporal decay of the linear polarization
for the resonant QD at $T = 77\,$K (upper curves) and $T =
300\,$K (lower curves). The excitation pulse is centered around
the lowest optical transition (h1-e1 exciton) with a spectral
window of $\pm 12\,$meV. The inset shows the initial decay.}
\end{center}
\end{figure}

The temporal decay of the polarization amplitude $|P(t)|$ is
displayed in Fig.\,\ref{Polaris} for the resonant QD. In the
calculation, we have taken into account only transitions which
are energetically close to the (lowest) h1-e1 exciton. Thus,
phonon satellites and the interference with the other transitions
are suppressed. This spectral window would correspond to a
finite-duration excitation pulse. For an elevated temperature of
300\,K, the initial decay goes much faster, which resembles a
traditional temperature-dependent dephasing. However, at larger
times, the polarization oscillates irregularly around a finite
value (therefore, we would like to use the term
\emph{quasi-dephasing}).

A similar dynamical decay due to incommensurate energies has been
discussed in Refs. \cite{K02,Jac03}. In real systems, these LO
phonon beats will be damped finally by anharmonicity
effects \cite{MJ05}.

For the absorption spectrum, we have chosen to broaden all
discrete lines in \Eq{DefAbsorption} with a fixed Gaussian of
variance $\sigma=1\,$meV. In this way, we can visualize both, the
transition energies and their oscillator strengths.
\begin{figure}[t]
\begin{center}
\includegraphics*[width=3in,angle=0]{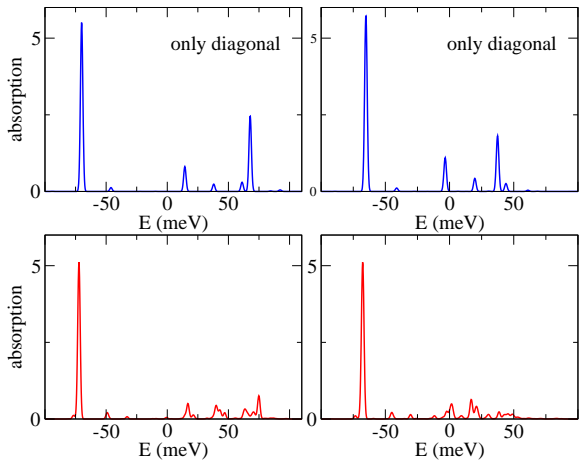}
\caption{\label{AbsorptionKF}Normalized absorption spectrum
$\alpha(\omega)$ for the non-resonant QD (left) and the resonant
QD (right) at $T=77\,$K. The lower panels refer to the full model
while the upper panels are restricted to level-diagonal phonon
coupling. The bare h1$\rightarrow$e1 transition energy is chosen
as zero of energy.}
\end{center}
\end{figure}

In Fig.\,\ref{AbsorptionKF}, the absorption spectrum
$\alpha(\omega)$ is shown for both the non-resonant QD (left) and
the resonant one (right). In the lower panels, all phonon
couplings are included. A set of closely spaced levels (bundle)
is seen around the two main optically allowed exciton transitions
h1-e1 and h2-e2. The excited-state bundle for the resonant QD
exhibits a particular broad range. A simple theory like e.g. the
second Born approximation \cite{KK98} would give here a real
broadening of the line, which may be interpreted as a real
dephasing process via energy-conserving phonon emission. We see
that in the present exact treatment, the situation is a bit more
complex, and can at best be called quasi-dephasing. Also, the
comparison with the non-resonant dot (left) shows that
quasi-dephasing is in no way restricted to exact resonance.
Although a treatment of exciton
occupation is outside the frame of the present paper, this
finding points to the absence of a clear phonon bottleneck in
polar quantum dots.

We have also calculated the spectrum for the reduced model of
level-diagonal phonon coupling $M_{\q ij}^a \propto \delta_{ij}$
(upper panels in Fig.\,\ref{AbsorptionKF}). This case is known to
be exactly solvable and called \emph{Independent Boson Model},
see Ref.\,\cite{Mahan}. We found excellent agreement supporting
our method, and could justify the chosen truncation in Boson
occupation. The level-diagonal absorption spectrum exhibits
phonon satellites which carry only a few percent of the total
weight, leaving the zero phonon line as the dominant feature. The
inclusion of the inter-level or non-adiabatic coupling thus
changes the absorption spectrum significantly which was already
emphasized in Ref. \cite{Fom98}.

\begin{figure}[t]
\begin{center}
\includegraphics*[width=3in,angle=0]{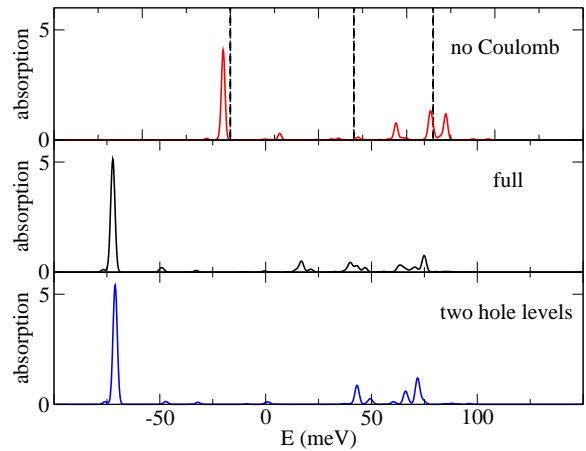}
\caption{\label{LevelEffects}Normalized absorption spectrum
$\alpha(\omega)$ for the non-resonant QD at $T = 77\,$K. Upper
panel: Without Coulomb interaction. The
transition energies without phonon interaction are shown as
vertical dashed lines which refer to the vertical arrows of Fig. \ref{Levels}. Middle panel: All couplings
and levels included. Lower panel: Only two hole levels included.}
\end{center}
\end{figure}

In Fig.\,\ref{LevelEffects} we show the different contributions
separately. Without Coulomb interaction (upper panel), the
transitions are slightly shifted compared to their bare positions
marked by dashed vertical lines. This allows to extract a polaron shift of the
lowest transition in the non-resonant QD of 4\,meV. This is much
smaller than the standard expression for the polaron shift in
\textit{bulk} CdSe (electrons: 11.3\,meV, holes: 21.1\,meV) and
shows indeed the strong charge cancellation in a QD as discussed
in Ref.\,\cite{SMC87}. In addition, the restriction to a finite
number of sublevels might underestimate our calculated QD polaron
shift. The comparison with the full calculation including phonon
and Coulomb interaction (middle panel) shows the importance of
the exciton effect. There is a general shift which is related to
the dominant diagonal Coulomb matrix element $v_{1111} = 67.4$meV.
However, the competition between nondiagonal Coulomb interaction
and phonon coupling (which are of the same order of magnitude)
modifies the spectrum in a complex manner, and cannot be reduced
to a rigid spectral shift. This clearly shows the importance of
excitonic effects in the absorption spectrum of strongly polar
quantum dots.

In the lower panel, results of a calculation with only two hole
levels are shown. The h3-e1 transition and its excitonic
modification is, of course, missing here. Nevertheless, the basic
features in the spectrum prevail. The polaron shift is a bit
smaller and the spectrum is less ``smeared out'' compared to the
full model with three hole levels (middle panel).

\subsection{Convolution approach}
Performing the unitary transformation described in  section
\ref{GramSchmidt} on the Hamiltonian for each band separately, we
end up with a model containing three Bosonic modes for the
conduction band and six Bosonic modes for the valence band. Using
the conservation of the electron (hole) number and again a
parabolic confinement potential, we finally have to diagonalize a
Hamiltonian with two Bosonic modes and two Fermionic states for
the conduction band and four Bosonic modes and three Fermionic
states for the valence band. For the two-state model, this has
been discussed at length in our previous paper
Ref.\,\cite{SZC00}.

Within the convolution approach, excitonic effects can only be
included approximately by a rigid shift of the spectrum using an
ad-hoc exciton binding energy. In the following discussion, we
want to demonstrate that even the electron-phonon interaction is
not treated appropriately. Therefore, we have no Coulomb
interaction in this subsection.

\begin{figure}[t]
\begin{center}
\includegraphics*[width=3in,angle=0]{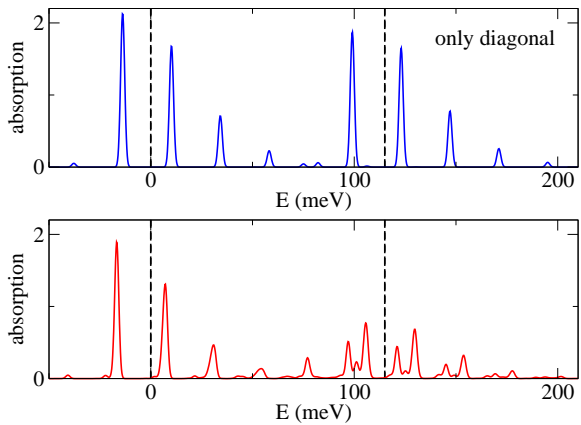}
\caption{\label{AbsorptionSF}Normalized absorption spectrum
$\alpha^{con}(\omega)$ obtained from convoluting the spectral
functions for electron and hole in the non-resonant QD. The
transition energies without phonon interaction are shown as
vertical dashed lines. The upper panel refers to the diagonal
phonon coupling, while the lower panel includes all coupling
matrix elements.}
\end{center}
\end{figure}

In Figure \ref{AbsorptionSF}, the absorption spectrum
$\alpha^{con}(\omega)$ from the convolution of the spectral
functions \Eq{AbsConvol} is shown for the non-resonant case. In
order to reach the same (artificial) broadening of the spectrum,
we have to broaden the spectral functions with a Gaussian of
reduced variance, $\sigma=1/\sqrt{2}\,$meV.

The inclusion of nondiagonal phonon matrix elements (lower panel)
is not so important here, as the spectrum is dominated by strong
phonon satellites. A comparison with the exact spectrum
(neglecting excitonic effects) of Fig.\,\ref{LevelEffects} (upper
panel) shows, however, that these satellites are much too strong.
As discussed at the beginning of Sec.\,\ref{SecAbsorption}, it is
the correlated phonon scattering between electron and hole states
which reduces the satellite structure appreciably. Thus, the
one-particle spectral functions are not able to describe the
optical transitions properly. Thus, we conclude that the
convolution approach as applied in Refs.\cite{KK98,VAM04} is
insufficient, even if the phonon coupling is of moderate
strength, as in the present example.
%
%
%
%

\section{Summary}
We have presented a solvable model which describes the optical
properties of a single quantum dot interacting with LO phonons,
and includes exciton formation. The linear polarization and the
corresponding absorption spectrum is calculated by diagonalizing
the appropriate Hilbert space of six electronic levels and six
Bosonic modes split off from the full phonon modes via an
orthogonalization procedure. The correct evaluation of the
two-particle dipole-dipole correlation function is contrasted
with a simplified approach using the spectral convolution of
one-particle spectral functions. The numerics shows this
convolution approach to be inadequate, as it gives phonon
satellites strongly enhanced compared to the correct result.
Qualitatively, this can be traced back to a missing compensation
between different contributions in a diagrammatic analysis.

Parameter values for CdSe quantum dots are used, and a comparison
between resonant and non-resonant QDs is carried out. Signatures
of the phonon bottleneck are found, i.e. a stronger
quasi-dephasing for the resonant case, but this is not as
dramatic as simple arguments using Fermi's golden rule for phonon
emission would predict.

\section{Acknowledgments}
Funding from MCyT (Spain) through grant MAT2002-04095-C02-01 and
from DFG (Germany) through Sfb 296 is acknowledged.
%
%
%
%
\begin{appendix}
\section{Electron-Hole transformation}\label{AppEHTrafo}
Switching from the conduction-valence-band description  to the
electron-hole picture produces a shift in the Bosonic operators.
Here we show that only the shift of the zero-momentum mode
survives.

To proceed we have to use the complete form of the confinement
wave  function including the Wannier basis $w_a({\bf r})$ of the
band under consideration,
\be \Psi_j^a(\r)=\sum_\R \psi_{j \R}^a \, w_a(\r-\R) \, ,\ee
where  $\psi_{j \R}^a$ is the envelope part (called confinement
function in the previous sections). The orthogonality of the
(real) wave functions induces
\be \delta_{jl} = \int_\Omega d^3 r \,\Psi_j^{a}(\r) \,
\Psi_l^a(\r) = \sum_{\bf R} \psi_{j \R}^{a} \, \psi_{l \R}^a \, ,
\ee
since the Wannier functions are orthogonal on different lattice
sites $\R \neq \R'$. Dropping prefactors, we evaluate
\bea \sum_j M_{\q jj}^{a} & \propto & \int_\Omega d^3 r\,
 \sum_{\R \R'j} \psi_{j \R}^{a} \,\psi_{j\R'}^a \, e^{i\q\r} \\
 &\times & w_a(\r-\R) \, w_a(\r-\R') \equiv C_\q \, . \nonumber \eea
The sum over $j$ gives a Kronecker symbol $\delta_{\R \R'}$ due to
the completeness of the coefficients $\psi_{j \R}^a$. The
remaining integral simplifies to
\bea C_\q & = & \int_\Omega d^3 r \sum_\R e^{i\q\r}
\, w_a^2(\r-\R) \\
 &= & \int_\Omega d^3 r'\, e^{i\q\r'} \, w_a^2(\r') \sum_\R
 e^{i\q\R} \, . \nonumber \eea
The last sum over $\R$ produces ${\cal N}\delta_{\q 0}$, where
${\cal N}$ is the number of elementary cells $\Omega_0$ within the
normalization volume, $\Omega={\cal N}\,\Omega_0$. Altogether we
have
\be C_\q = \delta_{\q 0} \, {\cal N} \, . \ee
Therefore, the correction in the last line of \Eq{HamEH} reduces
to zero momentum,
\bea \left(b_0\dag+b_0\right)\sqrt{{\cal N}\hbar\omega_0}\, \alpha_v(q)
\, , \\
\alpha_v(q)=\sqrt{\frac{1}{\Omega_0}\frac{e^2}{2q^2\varepsilon_0}
\left(\frac{1}{\varepsilon_\infty} -\frac{1}{\varepsilon_S}
\right)} \, . \nonumber \eea
In the prefactor $\alpha_v(q)$, we let $q\to 0$ at the end.

In order to remove the last line from the Hamiltonian, the
zero-momentum phonon operator is shifted according to
\be b_0 + \alpha_v(q)\sqrt{{\cal N}/\hbar \omega_0} \rightarrow
b_0 \, ,\ee
which brings from the free phonon Hamiltonian a quadratic
contribution ${\cal N} \alpha_v^2(q)$. Together with the valence
energy sum, this gives an unimportant $c$ number contribution
which can be dropped. However, the electron-phonon interaction is
getting an addition at $\q \to 0$, too. Due to the orthogonality
of the confinement functions, we have in leading order
\be M_{\q jl}^a = \sqrt{\frac{\hbar\omega_0}{\cal N}} \,
\alpha_v(q) \left(\delta_{jl} + O(q) \right) \, , \ee
which gives as correction
\be -2\alpha_v^2(q) \left(e_i\dag e_i \, - \, h_i\dag h_i\right)\,
. \ee
However, this term vanishes for any state having no or an equal
number of electrons and holes (charge neutrality). For optical
excitation and recombination, this is just the relevant sector of
the Hilbert space. Therefore, the singularity in
$\alpha_v(q)\propto 1/q$ (which would have to be treated
carefully) does not contribute to physical quantities.

Finally, we note that the above argument also holds for acoustic
phonons.
%
%
%
\section{Gram-Schmidt orthonormalization}\label{AppGramSchmidt}
We apply the Gram-Schmidt scheme by first constructing the
orthogonal operator set
\be \tilde{B}_1 = A_1 \; , \quad \Btl = A_\lambda \, - \,
\sum_{\alpha = 1}^{\lambda-1} \left[A_\lambda, B\dag_\alpha
\right] B_\alpha \, ,\ee
and normalizing it afterwards via
\be B_\lambda = N_\lambda \Btl \quad \mbox{where} \quad
N_\lambda^{-2} = \left[ \Btl, \Btl\dag\right] \, .\ee
The new operators obey the canonical commutation rules
\be \left[B_\lambda, B\dag_\alpha \right] = \delta_{\lambda
\alpha} \, .\ee
The final result can be written as a linear transformation
\be \label{BfromA} B_\lambda = N_\lambda \sum_{\alpha=1}^\lambda
K_{\lambda\alpha} A_\alpha \, .\ee
By construction, $K_{\lambda \alpha}$ is a tridiagonal matrix
which has nonzero elements for $\lambda \ge \alpha$ only. The key
ingredient for its evaluation is the commutator
\bea \label{Mlambdaalpha} \left[A_\lambda, A\dag_\alpha \right] &
= & \sum_{\q\,\q'} M_{\q \lambda} M_{\q'\alpha}^* \left[\bq,
b\dag_{\q'} \right] \nonumber \\
 &= &\sum_\q M_{\q \lambda} M_{\q \alpha}^*
\equiv M_{\lambda \alpha} \, , \eea
and we get a recursive determination according to
\bea K_{\lambda\lambda} & = &1 \, ,\\
K_{\lambda \alpha} &=& - \sum_{\beta=\alpha}^{\lambda-1} N_\beta^2
K_{\beta \alpha} \sum_{\nu=1}^\beta M_{\lambda \nu}K_{\beta \nu}^*
\quad (\lambda > \alpha)\, . \nonumber \eea
The norm follows from
\be N_\lambda^{-2} = \sum_{\alpha,\beta=1}^\lambda
K_{\lambda\alpha} M_{\alpha \beta} K_{\lambda\beta}^*  \, .\ee
In order to transform the Hamilton operator to the new phonon
operators $B_\lambda$, we need to invert \Eq{BfromA},
\be A_\lambda = \sum_{\alpha=1}^\lambda I_{\lambda\alpha} B_\alpha
\, .\ee
Again, $I$ is tridiagonal and can be determined quite easily via
\be I_{\lambda\lambda} = 1/N_\lambda \, , \quad \lambda
> \alpha: \, I_{\lambda \alpha} = - \sum_{\beta=\alpha}^{\lambda-1}
K_{\lambda\beta} I_{\beta \alpha} \, . \ee

%
\section{Coupling matrix elements}\label{AppCoupling}
The standard Fr\"ohlich coupling for the electron-LO-phonon
interaction is applied to the dot confinement states,
\bea \label{Froehlich} M_{\q jl}^{a} & = &
\sqrt{\frac{\hbar\omega_0}{\Omega}\frac{e^2}{2q^2\varepsilon_0}
 \left(\frac{1}{\varepsilon_\infty}
         -\frac{1}{\varepsilon_S}\right)} \,\Phi_{\q jl}^a \, ,\\
\Phi_{\q jl}^a& = & \int d^3r\, \psi_j^a({\mathbf{r}})\,
      e^{i\q\cdot\mathbf{r}}\,\psi_l^a({\mathbf{r}}) \quad
      (a=c,v) \nonumber \,.\eea
In the subspace of one electron-hole pair states, the Coulomb
interaction $H^C$ reduces to
\be \label{CoulombGen} \left<n_\lambda, kl | H^C | n'_{\lambda},
ij \right> = -\delta_{n_\lambda,n'_\lambda}\, v_{klij} \ee
which is diagonal in the phonon quantum numbers. The matrix
element is given by
\bea \label{CoulombME} v_{klij} &= & \int \!d^3r\, d^3r'
\psi_k^c(\r)\, \psi_l^v(\r')\, \frac{e^2}{4\pi\varepsilon_0
\varepsilon_S
|\r-\r'|}\,\psi_j^v(\r') \,\psi_i^c(\r) \nonumber\\
&=&\frac{1}{\Omega}\sum_\q \Phi_{\q ki}^c\frac{e^2}{\varepsilon_0
\varepsilon_Sq^2}\Phi_{-\q lj}^{v} \, .\eea
Diagonalizing the phonon-free part together with $H^C$ would lead
to (non-polar) exciton transition energies. However, in the
present context it is more appropriate (and easier) to
diagonalize phonon and excitonic effects together. Consequently,
the main interband transition energies are related to
exciton-polaron states in the quantum dot.

For simplicity we consider an anisotropic parabolic potential as
dot confinement for both, electrons and holes, with $x$ as the
long axis. The three energetically lowest wave functions are given
by
\bea \label{OsciEF} \psi_1^a(\mathbf{r})& = & \frac{1}{N_a}
\exp-\frac{1}{2}
  \left(\frac{x^2}{X_a^2} + \frac{y^2}{Y_a^2} +
  \frac{z^2}{Z_a^2} \right) \, ,
    \nonumber \\
 \psi_2^a(\mathbf{r})& = &\sqrt{2}\frac{x}{X_a}\psi_1^a(\mathbf{r}) \, ,\\
 \psi_3^a(\mathbf{r})& = &
    \left(\sqrt{2}\frac{x^2}{X_a^2}-\frac{1}{\sqrt{2}}\right)
    \psi_1^a(\mathbf{r})\; \nonumber ,
\eea
where $X_a>Y_a, Z_a$ are the spatial extensions (variances) of
the ground state, and $N_a^2=\pi^{3/2} X_a Y_a Z_a$ its
normalization.

The matrix elements in \Eq{Froehlich} and \Eq{CoulombME} thus
read
\bea \Phi_{\q 11}^a & = &\exp \left(-\frac{1}{4}\left(q_x^2 X_a^2
+ q_y^2 Y_a^2 + q_z^2 Z_a^2\right)\right)  \, ,\nonumber \\
\Phi_{\q 21}^a & = & iq_x X_a/\sqrt{2} \,\Phi_{\q 11}^a \, ,\nonumber \\
\Phi_{\q 31}^a & = & -q_x^2 X_a^2/\sqrt{8} \,\Phi_{\q 11}^a \, ,\nonumber \\
\Phi_{\q 22}^a & = &
    \left(1 - q_x^2 X_a^2/2\right)\Phi_{\q 11}^a\, ,\\
\Phi_{\q 32}^a & = &
    i\left(q_x X_a - q_x^3 X_a^3/4\right)\Phi_{\q 11}^a\, ,\nonumber \\
\Phi_{\q 33}^a & = &
    \left(1-q_x^2 X_a^2 + q_x^4 X_a^4/8\right)\Phi_{\q 11}^a
\; . \nonumber
\eea
Notice that $\Phi_{\q 22}^v = \Phi_{\q 11}^v + \Phi_{\q
31}^v/\sqrt{2}$. Therefore, one additional Bosonic mode can be
eliminated.

The new coupling constants $M_{\lambda\alpha}$ are obtained by
integrating a pair of coupling constants $M_{\q ij}^a$ over $\q$,
\Eq{Mlambdaalpha}. This final integration can be reduced to the
following (elliptic) integrals,
\be \label{In} J_n = \int_1^{\infty} dt \, \frac{t^{-n}}
  {\sqrt{t(t-p)(t-q)}} \quad (n = 0, 1, 2, 3, 4)  \ee
with $p = 1-(Y_a^2+Y_b^2)/(X_a^2+X_b^2)<1$ and $q =
1-(Z_a^2+Z_b^2)/(X_a^2+X_b^2)<1$. For a cylindrically symmetric
potential with $Y_a=Z_a$, we have $p=q$, and the integrals reduce
to simple analytic functions,
\be  J_0 = \frac{1}{\sqrt{p}} \log\frac{1+\sqrt{p}}{1-\sqrt{p}}\,
, \quad J_{n+1} = \frac{J_n - (n+1/2)^{-1}}{p} \, .\ee
The same type of integration over $\q$ appears in the Coulomb
matrix elements \Eq{CoulombME}. For example, the diagonal matrix
element for the lowest transition h1-e1 can be evaluated as
\be v_{1111} = \frac{e^2}{ \varepsilon_0 \varepsilon_S \sqrt{X_c^2
+X_v^2}}\, \frac{J_{n=0}}{4 \pi^{3/2}} \ee
where $p$ and $q$ in \Eq{In} are determined with $a=c, \, b=v$.
%
%
%
%
\section{Material parameters}\label{AppParameters}

Parameter values for the polar semiconductor material CdSe
forming the quantum dot are from Ref.\,\cite{Puls99} as LO-phonon
energy $\hbar\omega_0=24\,$meV, conduction band mass
$m_c=0.13\,m_0$, and valence band mass $m_v=0.45\,m_0$.
Dielectric constants have been taken from Ref.\,\cite{LanBoer}:
$\varepsilon_S = 9.57$, $\varepsilon_\infty = 6.27$.

We assume that the confinement potentials of electrons and holes
are scaled by a fixed ratio $\rho$ which equals the ratio between
conduction band offset and valence band offset for barrier (ZnSe)
and dot material (CdSe), $\rho = 5.44$ \cite{Puls99}. Applying
this to the parabolic potential, we find a unique ratio between
the lengths in the oscillator eigenfunctions \Eq{OsciEF},
\be \label{offset} \xi\equiv\frac{L_c}{L_v} =
\left(\frac{m_v}{m_c\rho}\right)^{1/4} \, , \quad (L = X, Y, Z)\ee
which equals $\xi = 0.89$.

Since the $x$ direction is taken as the longest one, the
energetic distance between ground level ($j=1$) and first excited
level ($j=2$) is exclusively given by the $x$-confinement,
\be  \label{Leveldist} \Delta^a\equiv|\epsilon_2^a -
\epsilon_1^a|  =  \frac{\hbar^2}{m_a X_a^{2}} \quad (a = c, v) \,
. \ee
Taking experimental values which have been reported for a certain
CdSe quantum dot in Refs.\,\cite{FlissHen01,Akimov04} ($\Delta^c=
80\,$meV, $\Delta^v = 35\,$meV) we obtain according to
\Eq{Leveldist} $X_c=2.71\,$nm and $X_v = 2.20\,$nm. This yields a
length ratio of $\xi=X_c/X_v=1.23$ which is larger than the value
derived from \Eq{offset}. However, a possible alloying of the QD
material \cite{GKFNZGT04} and deviations from the (idealized)
parabolic confinement potential may introduce substantial
uncertainties. In the numerical calculations, we have used $\xi =
1.23$ throughout. 

There is no direct experimental access to the
size of the QD in the other (shorter) directions. To keep things
simple, we take the same representative value for the two smaller
lengths: $Y_v=Z_v=1.3\,$nm and $Y_c=Z_c=\xi Y_v=1.6\,$nm. The
next levels corresponding to this shorter length would have a
spacing of $100\,$meV in the valence band. Since this is larger
than twice the level spacing corresponding to the larger $X$
extension ($70\,$meV), we have properly selected the lowest hole
confinement states in \Eq{OsciEF}.

It is easily seen from the parity of the oscillator wave
functions \Eq{OsciEF}, that the optical transitions
h1$\rightarrow$e2, h2$\rightarrow$e1 and h3$\rightarrow$e2 are
forbidden.  The remaining non-zero dipole matrix elements
\Eq{DipoleDef} can be expressed by the same length ratio $\xi$
introduced above. In units of $\mu_{cv}$, we find
\bea \mu_{11}  &=&  \left(\frac{2\xi}{1+\xi^2}\right)^{1/2} \, ,
\quad \mu_{22} = \left(\frac{2\xi}{1+\xi^2}\right)^{3/2} \, ,\nonumber\\
\mu_{31}&=& -\left(\frac{\xi}{1+\xi^2}\right)^{1/2}
\frac{1-\xi^2}{1+\xi^2}\,. \eea
With $\xi=1.23$, we get $\mu_{11}= 0.99$, $\mu_{22}=0.97$ and
$\mu_{31}=0.14$. Such values very close to a Kronecker delta are
typical for the strong confinement in a quantum dot. The
energetic distance between the dominant transitions equals
$\Delta^c + \Delta^v = 115\,$meV (without polaron and Coulomb
corrections).

Apart from the parameter set deduced from the experimentally
given level distances, we chose another one which has perfect
resonance between the hole level distance and the LO phonon
energy, i.e. $\Delta^v=24\,$meV. This yields a somewhat larger QD
with $X_v=2.66\,$nm and $X_c=\xi X_v=3.27\,$nm, while keeping all
the other parameters the same. The level distance for the
electrons amounts to be $\Delta^c=55\,$meV, giving an energetic
distance between the main allowed transitions of 79\,meV. We call
this the \emph{resonant} QD, while the parameter set backed by
experiment is referred to as the \emph{non-resonant} QD.

The diagonal Coulomb matrix elements are in the range of
50-70\,meV which is four times the exciton binding energy in bulk
CdSe (15\,meV) and quantifies the strong enhancement of Coulomb
effects in small QDs. The non-diagonal Coulomb matrix elements
are around 15\,meV which is comparable to the effective phonon
coupling strengths.
\end{appendix}

\end{document}